# Investigating Lipid Bilayer Self-Assembly and Formation of Ripple Phase: Insights from a Coarse-Grained Implicit Solvent Model

Biplab Bawali[*], Alokmay Datta, Jayashree Saha

*Department of Physics, University of Calcutta, 92 Acharya Prafulla Chandra Ray Road, Kolkata 700 009, India*

*Corresponding author: bbphy_rs@caluniv.ac.in

In this study, we present a comprehensive exploration of formation of different phases in lipid molecules using a coarse-grained implicit solvent model, where each lipid molecule is represented as a rigid, three-bead rod-like structure. Our study not only successfully replicates the spontaneous self-assembly of lipid bilayers but also elucidates the intricate phase transitions between the gel phase, liquid phase, and the elusive ripple phase $(P_\beta)$. Specifically, we uncover the pivotal role of molecular rigidity in promoting the formation of the ripple phase. The significance of our findings lies in their potential to reshape our understanding of lipid bilayer dynamics and phase transitions. By shedding light on the ripple phase, a phase previously difficult to simulate convincingly, the insights gained from this study have the potential to guide future investigations into the behavior of biological membranes and their role in cellular processes.

## I. INTRODUCTION

The cell membrane separates the interior of a cell from the exterior region. It is essentially a phospholipid bilayer embedded with cholesterol, proteins, carbohydrates etc. [1]. Phospholipids are mainly responsible for giving rise to the basic bilayer matrix in the cell membrane. A phospholipid molecule consists of a hydrophilic head and two hydrophobic tails. The hydrophilic part has a $PO_4^-$ and a $NH_3^+$ group, which makes it polar and this is why the head group interacts with water molecules. On the other hand the hydrophobic part of this molecule consists of two chains with 16 C-atoms each. The various biological roles of the membrane lipids are related to the different phases arising out of the competition between mutual interactions of heads and tails and their interactions with water and other biologically relevant molecules. The lipid membrane, which has molecule-specific permeability and regulates how chemicals enter and exit cells, is essential for maintaining cell potential [2]. The relative mobility (fluidity) of individual lipid molecules, as well as how this mobility varies with temperature, is one of the essential aspects of a lipid bilayer. This response is known as the phase behaviour of the bilayer. These bilayers can have various fluid and solid phases depending on the interaction of one lipid molecule with its neighbours [3] mainly at different temperature and solvent concentration.

The phospholipid bilayer is found in $L_\beta$ 'gel' phase at lower temperatures. At higher temperatures, the bilayer transits to a 'fluid' $L_\alpha$ phase, due to increased mobility of individual lipids in the bilayer. Between these two phases a smectic ripple phase, $P_\beta$, is detected in hydrated lipid bilayers. The long range corrugations in the layers of the membrane with a well-defined periodicity along an axis parallel to the mean bilayer plane characterise this ripple phase. Usually when the ratio of head group cross-sectional area to the total cross-sectional area of tails surpasses a particular value, ripples emerge to alleviate packing frustrations [4]. Therefore, not all lipids show the ripple phase at their bilayer configuration. Phospholipids, which have phosphatidylcholines (PC) as head groups and were extensively investigated, fall in the group of ripple-forming lipids. Their phase behaviour, in particular this ripple phase, is of considerable importance to biological system.

The ripple phase frequently coexists with the liquid and gel phases, with the relative stability of the three phases dependent on the temperature, lipid composition, and other variables. The ripple phase gets more stable as the temperature drops until eventually it is the only stable phase. The bilayer enters the gel phase at extremely low temperatures [5–7].

Till date, in spite of the availability of sufficient experimental and simulation data [5], there is yet a lack of molecular level understanding of the structural changes in lipid membranes in intra- and extra-cellular fluids. Tardieu et al. [8] first published the result of an extensive x-ray diffraction study of lipid phases including the ripple phase in a variety of lecithin-water systems prepared from hen eggs. They came to the conclusion that, for fully extended chains, the orientation is only constrained to the lamellae's normal if the chains are chemically heterogeneous, but, for homogeneous chains, the orientation tilts and the lamellae are deformed by periodic ripples. A periodic sinusoidal structure of the ripple phase $P_\beta$ was reported by Tardieu from the electron density distribution profile. At least two distinct ripple states have been observed by Tenchov et al. [9] and they frequently coexist. Later in 2003, Sengupta et al. [10] reported electron density maps (EDMs) for the ripple phases that were formed by several phosphorylcholine lipids. While DPPC has a metastable sinusoidal symmetric ripple phase, lipids like DMPC, POPC, DHPC, and DLPC exhibit asymmetric ripples that lack a mirror plane normal to the ripple wave vector. For enhanced clarity, we have included a schematic diagram in Fig. 1 depicting the symmetric and asymmetric ripple phases [10–13].

The Dissipative-Particle Dynamics (DPD) simulation of a coarse-grained lipid model developed by Kranenburg et al. [14] was the first to successfully reproduce a periodically modulated membrane state. They looked at how temperature, head group interaction, and tail length affected the phase behaviour of double tail lipids. They found that the head groups must be sufficiently encircled by water in order for the rippling structure to form. This requirement was attained by raising the head-head repulsive parameters. As a result they were able to create the periodic ripple phase for lipid molecules, which consists of two hydrophobic tails with lengths ranging from 4 to 7 beads and a head group made up of three hydrophilic beads. The rippled structure is neither asymmetric nor does it exhibit the sinusoidal symmetric ripple phase [10–12,15], and the distribution of head groups in that structure did not match the earlier experimental findings.

In the same year, de Vries et al [16] conducted another all-atom molecular dynamics simulation on a lecithin bilayer. For the DPPC molecule, they have found an asymmetric ripple structure with two domains, M and m that differ in length and bilayer thickness. The two monolayers are separated from one another and tilted relative to the bilayer normal in the M domain whereas the lipids tails of the two monolayers are completely inter digitated in the m domain and the molecules in the concave part of the kink are highly disordered. As of now, this was the best illustration of an asymmetric ripple structure, but this result conflicted with result of Sengupta et al. [10]that DPPC molecules display symmetric ripple phase only.

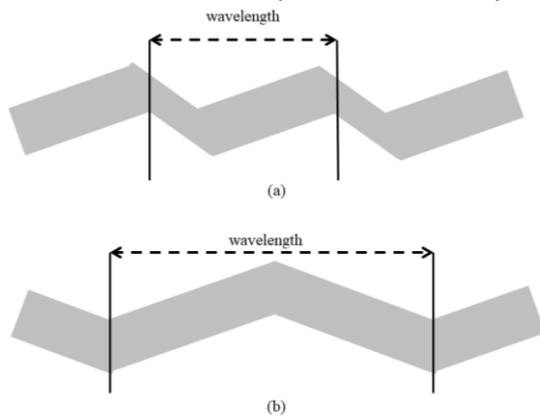

FIG. 1. Schematic diagram of - (a) Asymmetric ripple structure; (b) Symmetric ripple structure

In 2007, Lenz and Schmitt [11] published a Monte Carlo simulation result of a simplified coarse-grained lipid model that effectively recreated both symmetric and asymmetric ripple states. The asymmetric ripple's structure matched de Vries [16] theory rather well. However, this time it had a periodic structure and it was also generic. A chain with one head bead and six tail beads that is joined by anharmonic springs serves as the paradigm for the representation of lipid molecules. In 2008, Sun and Gezelter [17] proposed an explicit solvent model taking 10 water molecules per single lipid considering a rigid molecule with a dipolar spherical head and an ellipsoidal tail and achieved ripple phases for the bilayer using a molecular dynamics simulation. The dipolar interaction was the key for maintaining the bilayer structure exhibited by this particular model. This result point to both the redundancy of the flexible model and the importance of the lipid-water interaction and provide motivation for an implicit solvent model with a rigid molecule with fewer beads, to capture the physical essence of the phase behaviour of the lipid system. In 2012, Huang et al. [12]did this type of study using a four beads flexible model. The coarse-grained model for lipid bilayer self-assembly and dynamics, as described by Huang et al. [12], follows a mesoscopic strategy that blends molecular dynamics simulations for lipids with multiparticle collision dynamics for the solvent. In this model, the solvent is explicitly included using multiparticle collision dynamics which is usually implemented to preserve momentum. They successfully achieved the self-assembly of the bilayer and were able to reproduce the gel phase at the lower temperature and the liquid phase at the higher temperature. However, they missed the ripple phase of bilayer between these two phases. Several experiments [18,19] and coarse-grained [20–26], all atom [25,27–31] molecular dynamics simulations have been carried out by many researchers for the last 10 years. However, none of them could emphasize the role of molecular rigidity in ripple phase formation.

All-atom takes up huge computer runtime, and is necessarily limited to small system sizes and short time scales. Both are insufficient to mimic the huge spatiotemporal scales involved in cellular processes, such as self-assembly of vesicles or structural instabilities of membranes [12]. Again an all-atom simulation may 'overkill' since the most non-invasive experimental studies may not resolve the position and/or motion of individual atoms in these complex systems but only provide information about averaged structures. By neglecting some of the atomistic degrees of freedom but by incorporating the relevant interactions prudently, a coarse-grained (CG) model allows us to overcome both the spatial and temporal limitations of all atom models [32]. This has led to the development of CG simulation methods for lipid membranes, where a lipid molecule is modelled as a rod-like body consisting of some spherical beads. One of them represents the head group and others make up the tail part of the lipid molecule [33]. In this model all the individual interatomic forces are replaced by head-head, head-tail and tail-tail interactions through realistic coarse-grained interaction potentials.

Sun and Gazelter [17] and Huang et al.[12] both have used explicit solvent model but for a system large enough to represent a phase, these simulations require a large computing time. It is to be noted that solvent molecules, in this case water, occupy about 90% of the system, and simulation time is mostly consumed not only by the lipid-solvent interaction but also the largely redundant interaction between solvent molecules. The best way to reduce the computation time is to get rid of the solvent and consider the solvent interaction implicitly. The solvent is then represented by its effect on the head-tail, head-head, and tail-tail interactions, i.e. the interactions are modified to account for the solvent [34]. As a result the number of interacting particles and degrees of freedom is hugely reduced [35]. This is known as a free or solvent free or implicit solvent model [36].

Cooke and Deserno [34] presented an implicit solvent coarse-grained (ISCG) model to simulate lipid membranes. In this model the coarse-grained structure of a lipid molecule is exactly similar to the structure proposed by Farago [37] except the lipids in Farago model were rigid but here Cooke and Deserno used flexible molecules but, due to the presence of sufficiently broad attractive potential of the tails, self-assembly into stable bilayer is accomplished. Also, the two body potential is highly tuneable in this model [34]. However, the complexity with this model is that the molecules are not rigid. The three beads are linked by two finite extensible nonlinear elastic (FENE) bonds, which make the system complicated by bringing in more variable parameters that maybe not required for producing lipid bilayer, also that could not produce ripple phase.

Our aim is to evaluate the role of molecular flexibility, which results from the higher probability of the all-trans state than the gauche state. The importance of the flexibility cannot be verified directly from experimental result, so simulation study is needed. Our model molecules are governed by relatively simple interaction but they must have the following features: (i) they self-assemble to a fluid phase and (ii) the phases reproduce most of the known features of fluid membrane phases (gel, ripple, fluid) that can be verified experimentally.

We have modified the Cooke and Deserno [34] model by eliminating the flexibility of the molecule. Using this minimal model and performing Monte Carlo simulation under constant number, volume and temperature (NVT) conditions, self-assembly through spontaneous ordering from an isotropic phase to the fluid bilayer is achieved ergodically. Our coarse-grained model is also able to produce all the three phases during the increment in temperature, in particular the rippled phase is successfully reproduced. Our model shows explicit dipolar interaction between lipid and solvent molecules which takes huge amount of computational time[17] are not indispensable for bringing ripple phases.

## II. MODEL AND SIMULATION DETAILS

### A. MODEL

The schematic of the model is shown in the cartoon of Fig. 2. Each lipid molecule consists of three spherical atoms rigidly connected to each other. Atom1 represents the hydrophilic head group with added implicit solvent interaction, whereas atom 2 and atom 3 represent hydrophobic tail of the lipid molecule. The interactions are summarized schematically in the cartoons of Fig. 3. The different segments of the lipids interact with each other via Weeks-Chandler-Andersen (WCA) repulsive potential [38]which is the Lennard-Jones (LJ) potential, shifted upwards by $\varepsilon$ and truncated at the LJ potential minimum of $(2)^{1/6}\sigma$ [39].

$$\phi(r;\sigma_{ij}) = \begin{cases} 4\varepsilon\left(\left(\frac{\sigma_{ij}}{r}\right)^{12} - \left(\frac{\sigma_{ij}}{r}\right)^{6}\right) + \varepsilon, & r \leq r_c \\ 0, & r > r_c \end{cases} \quad (1)$$

With $r_c = (2)^{1/6}\sigma$, where $\sigma_{ij}$ is the effective diameter of each atom (Fig. 3(a)). For an effective cylindrical lipid shape $\sigma_{ij}$ is chosen as $\sigma_{head,head} = 0.95\sigma$, $\sigma_{head,tail} = 0.975\sigma$ and $\sigma_{tail,tail} = \sigma$ using Lorentz mixing rule [40] where $\sigma$ is the unit of length and $\varepsilon$ is the unit of energy. [34]

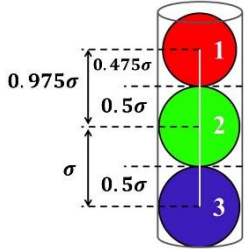

FIG. 2. The Three-atom Lipid Model

This stabilizing potential is more efficient than LJ potential due to its short range, as a smaller number of neighbours are required. The aqueous environment is simulated by an attractive (lipophilic) extra potential between all tail beads to model the hydrophobic interaction (Fig. 3(b)),

$$V_{tail}(r) = \begin{cases} -\varepsilon & , r < r_c + w_f \\ 4\varepsilon\left[\left(\frac{\sigma_{ij}}{r-w_f}\right)^{12} - \left(\frac{\sigma_{ij}}{r-w_f}\right)^{6}\right] & , r_c \leq r \leq w_f + w_{cut} \\ 0 & , r > w_f + w_{cut} \end{cases} \quad (2)$$

Here the range of the LJ potential is extended by inserting a flat length, $w_f$, which is the key tuning parameter. That is, tuning $w_f$ achieves a broad potential $V_{tail}$. The value of $w_{cut} =$ $2.5\sigma$ and $w_f$ is chosen as $w_f = 0.4\sigma$ following standard literature values.

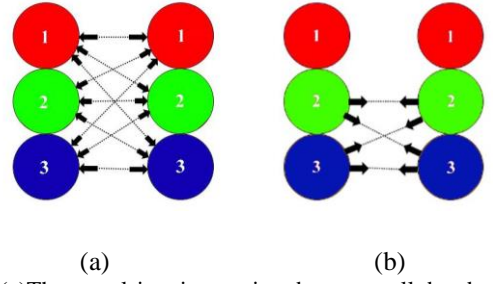

FIG. 3. (a)The repulsive interaction between all beads; (b) the additional attractive potential among tail beads only.

Therefore, head groupsof molecule I and molecule II are interacting only through the repulsive potential of equation 1 and this is also the interaction between the head and tail groups.Only the tail group of molecule I and the tail group of molecule II have the attractive potential of equation 2 in addition to the repulsive potential of equation 1.

### B. SIMULATION DETAILS

We have simulated two system of 800 and 1250 lipid molecules named as system-1 and system-2 respectively within two different cubic boxes with periodic boundary condition in all the three dimensions. Each molecule can make two moves in a single Monte-Carlo step, one is translation of the centre of mass and the other one is Euler rotation of the molecule around the centre of mass. The simulations were carried out at constant NVT conditions.

To get the right density we used trial-and-error method and finally box sizes $L = 21.8\sigma$ and $L = 27.3\sigma$ were chosen and reduced units ($\varepsilon = 1$, $\sigma = 1$) were used throughout. Minimum reduced pair separation between two atoms was set to $0.7\sigma$ and the reduced maximum displacement of each molecule in each step was $0.05\sigma$.

## III. RESULTS AND DISCUSSIONS

We first achieved the isotropic phase configuration of the system from a bilayer constructed manually. We have started from a bilayer configuration where all molecules were symmetrically placed in two parallel planes each plane containing half the total number of molecules. After 5000 MC steps at a sufficiently higher temperature of about $T^* = 5.0$ we got a configuration which is both orientationally and positionally isotropic as shown in Fig. 4. We considered this isotropic configuration as our initial configuration for both of our systems.

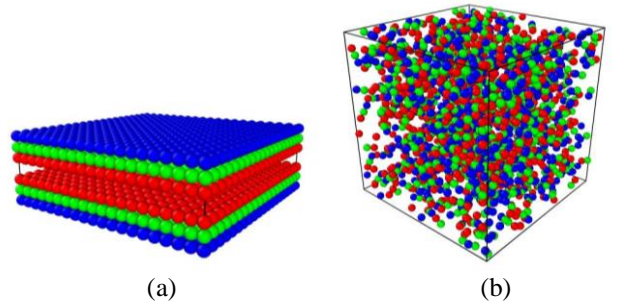

FIG. 4. (a) The initial pre-assembled configuration of 800 molecules; (b) the isotropic configuration at $T^* = 5.0$ after 5000 Monte Carlo steps.

We started lowering the temperature of the system until it reached the reduced temperature $T^* = 1.0$, corresponding to ambient temperature. We then kept the temperature fixed and simulated the system further with larger Monte-Carlo steps under identical conditions and without changing any other parameter. Both the systems reduced to a lamellar or gel($L_\beta$) phase after 8 hundred thousand MC steps. Snapshots of the configuration have been attached in Fig. 5.

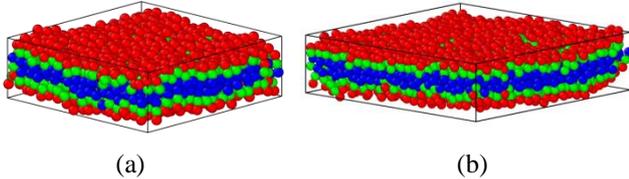

(a) (b)

FIG. 5. The lamellar gel phase achieved after 800k Monte Carlo steps at temperature $T^* = 1.0$; (a) system-1 of 800 molecules (b) system-2 of 1250 molecules

We followed this by starting to observe both the bilayers both on increasing and decreasing the system temperature. On reducing the temperature to $T^* = 0.6$ we got the solid phase whereas on increasing the temperature we obtained the symmetric ripple phase($P_\beta$) followed by the fluid phase $(L_\alpha)$ at $T^* = 1.2$ and $T^* = 1.375$ respectively in system 1 and at $T^* = 1.25$ and $T^* = 1.465$ respectively in system 2. Our focus is on discrete temperature observations that allow us to characterize the phases of the system qualitatively at suitable temperature —whether it resides in the gel, ripple, or fluid phase.

Sengupta et al. [10] confirmed through experimental studies that the ripple phases of DMPC, POPC, DHPC, and DLPC systems are all very similar in shape, and they all have the asymmetric ripple phase while molecule like DPPC exhibits a symmetric ripple phase $P_\beta$ on cooling from the $L_\alpha$ phase [10]. Lenz et al. [11]got the sinusoidal structure of the membrane on the symmetric ripple phase at $T^* = 1.18$ through MC studies very similar to our result.

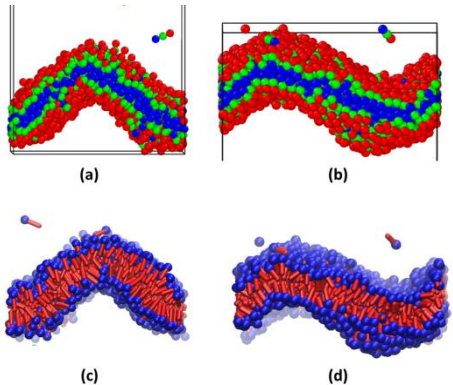

FIG. 6. The ripple phase observed at a higher reduced temperature of 1.2 in system-1 and at 1.25 in system-2. (a) ripple phase at temperature 1.2 in 800 molecule system, (b) ripple phase at temperature 1.25 in 1250 molecule system, (c) stick representation of ripple structure in 800 molecule system, (d) stick representation of ripple structure in 1250 molecule system

Since our model leads to the symmetric ripple phase it is more relevant to DPPC like molecules. As this is an implicit solvent model, the free space inside the box represents water. Comparison with the well-known lipid phase diagram [41], which depicts the effect of water content on lipid phases shows that the lipid bilayer shifts from $L_\beta \rightarrow P_\beta \rightarrow L_\alpha$ with increase in water concentration. Under acute dehydration, the system shows a richer phase diagram. Our system consists of 82% water. According to the phase diagram given by Carlson et al. [4]with increasing temperature a lipid bilayer under this condition of hydration has a transition from gel to fluid phase through a co-existing ripple phase.

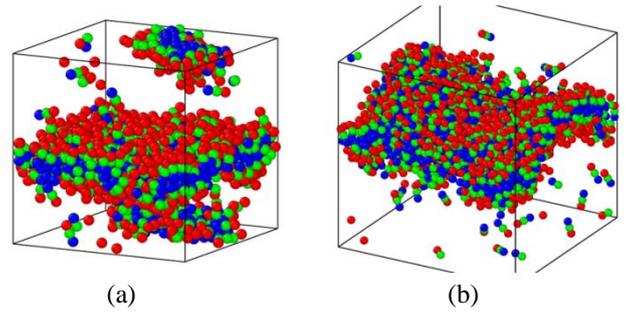

(a) (b)

FIG. 7. The fluid phase observed at a higher reduced temperature of 1.375 in system-1 and at 1.465 in system-2. (a) fluid phase at temperature 1.375 in 800 molecule system, (b) fluid phase at temperature 1.465 in 1250 molecule system

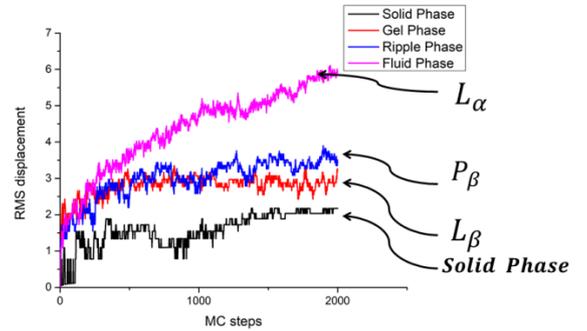

FIG. 8. Plot of rms displacement of molecules in different phases

Hence in this study we have reproduced this sequence with a minimal model with the snapshots of the phases displayed in Fig. 5-7. The rms displacement of lipid molecules in these phases has been plotted in Fig. 8 which confirms that the ripple is a coexistence phase between the $L_\beta$ and $L_\alpha$ phase [5–7,42]. This plot also supports that more thermal mobility occurs in the liquid $L_\alpha$ phase than gel phase $L_\beta$.

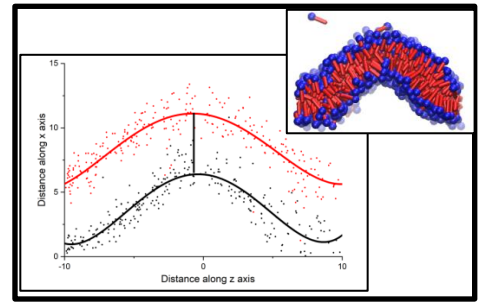

FIG. 9. Average width of the bilayer (separation between the head beads of upper and lower leaflets). A stick representation to understand the orientation of the molecules

At ambient temperature, head groups in the liquid phase have a higher area per molecule than those in the gel phase due to their loose packing and rotational flexibility.

The average thickness of the bilayer in system-1 is shown in the Fig. 9 where the overall shape of the bilayers is similar to that obtained from the EDMs of Sengupta et al. The peak-to-peak amplitude (shown in inset figure of Fig. 9) is about 46 Å and the ripple wavelength is about 204Å. Meanwhile, in System 2, the observed ripple phase at a temperature of 1.25 features a wavelength measuring 269 Å. Figure 10 illustrates the comparison between the ripple phases obtained in both systems. Table I displays the dependence of wavelength and amplitude, on temperature for the two different systems.

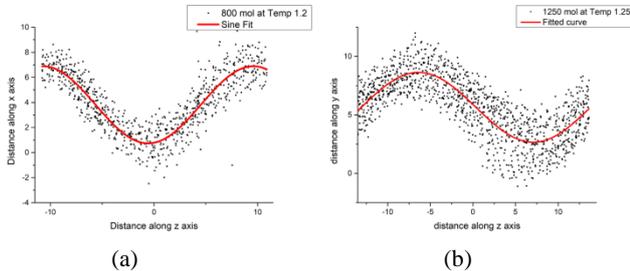

FIG. 10. Wavelength comparison of two system at the ripple phase; (a) the wavelength of ripple structure at temperature 1.2 in system 1 is 203 Å, (b) the wavelength of ripple structure at temperature 1.25 in system 2 is 269 Å

However, experimentally the symmetric ripple phase can only be found when the system is being cooled, whereas we are getting this phase while increasing the temperature of the system. Again, the asymmetric ripple phase is found in lipid molecules like DMPC, POPC, DHPC, with large mismatch between head and tail areas. This indicates that we can use our minimal model to mimic these different types of lipid molecules by altering the head tail size ratio, and thus may also achieve the asymmetric ripple phase. This work is underway. The absence of flexibility in our model and its success in explaining the phase space of hydrated bilayer points to the role of rigidity of the lipid molecule in bilayer assemblies. Hence, the construction of lipid bilayers and the development of the ripple phase depend critically on the rigidity of molecules [43–45].

TABLE I. The variation of Wavelength ($\lambda$) and Amplitude (A) with Temperature ($T^*$) at different system. $\lambda$ and A are in angstrom (Å) while the temperature is in LJ reduced unit. $\delta\_1$ and $\delta\_2$ are the possible standard deviation corresponding to $\lambda$ and A respectively

|  | ($T^*$) | $\lambda$ | $\delta\_1$ | A | $\delta\_2$ |
|---|---|---|---|---|---|
|  | 1.10 | 222.18 | 0.26366 | 13.7746 | 0.04308 |
|  | 1.15 | 211.18 | 0.13884 | 25.3079 | 0.05293 |
| System-1 | 1.2 | 203.08 | 0.19616 | 30.7433 | 0.06307 |
| (800 mol.) | 1.22 | 191.74 | 0.19018 | 30.4035 | 0.07853 |
|  | 1.25 | 199.12 | 0.22898 | 29.4626 | 0.09252 |
|  | 1.10 | 256.33 | 0.34347 | 9.3385 | 0.05481 |
|  | 1.15 | 280.06 | 0.17065 | 19.1256 | 0.04241 |
| System-2 | 1.25 | 268.96 | 0.15256 | 29.8728 | 0.07279 |
| (1250 mol.) | 1.30 | 269.29 | 0.16175 | 37.708 | 0.08162 |
|  | 1.35 | 246.64 | 0.20081 | 28.354 | 0.09335 |

The observed wavelength range of the ripple phase by Kaasgaard et al.[46] aligns with our findings, despite their experiment focusing on a double bilayer of DPPC lipid on a mica substrate. Direct comparison with experimental studies is challenging due to their investigation of deposited bilayers over solid substrates like mica or silica, distinct from our study's focus on bilayers immersed in water as this happens in physiological system. To derive ideal results, simulations become imperative, allowing for a more comprehensive understanding of the experimental result in a qualitative way. Analytical studies in this field are rare; theoretical studies involving mean-field approximation (ref. mercilia 2nd paper) were incapable of handling complex interaction for these organic molecules and their phase transitional properties. The potential used in this study has been widely used by the community engaged in lipid bilayer simulation however our restriction on molecular flexibility can be justified because the phases we have studied are relatively low temperature phases therefore all trans phase are more probable. Also, one chain of the lipid helps mutual ordering of the second chain bringing more rigidity to the lipid molecule.

Still we have verified our result with some renowned experimental observation by Sengupta et al,.[10] and Kaasgaard et al.,[46] along with some famous computational observation by De Vries et al.,[16] Sun and Gezelter[17], Lenz and Schmid[11] and Huang et al,.[12]. Apart from that our model is able to reproduce different phases of lipid bilayer with temperature successfully.

Our simulations account for the temperature-dependent variations in the ripple phase wavelength, a crucial insight validated within our model. Hence, our model offers a new perspective on the intrinsic characteristics of lipid molecules conducive to ripple phase formation. Remarkably, our study stands as the singular model demonstrating the self-assembly of lipid bilayers preceding the attainment of a ripple phase, amidst other discernible phases.

## IV. CONCLUSION

We have introduced a coarse-grained models for lipid molecule specifically for molecules like DPPC. In this Monte Carlo simulation study considering a 3-bead implicit solvent coarse-grained, i.e., a minimal model potential for rigid elongated lipid molecules, we successfully realized a stable, ordered, fluid bilayer structure ergodically, starting from an isotropic configuration. As molecular flexibility does not play major role in the ordered bilayer phase occurring at lower temperature region, this feature has been neglected in this model, making it significantly computationally inexpensive but capable of generating self-assembled lipid bilayers. Apart from self-assembly this model is also able to produce all the three phases i.e. gel phase, ripple phase and liquid phase of lipid bilayer with the variation in temperature. To the best of our knowledge the present work is able to reproduce the ripple phase along with transition in $L_\beta$ and $L_\alpha$ at suitable temperature and density for the first time.

The only drawback of this model is that we got only the symmetric ripple phase and an extension of the model with suitable modification, to the asymmetric ripple phase is underway.

## V. ACKNOWLEDGEMENT

BB gratefully acknowledge the support of Council of Scientific & Industrial Research (CSIR), India, for providing Senior Research Fellowship (File no: 09/028(1073)/2018-EMR-I )